\documentclass[12pt]{article}
\usepackage{graphicx}
\usepackage{epsfig}
\usepackage[utf8x]{inputenc}
\usepackage[T2A]{fontenc}
\usepackage[english]{babel}
\usepackage{textcomp}
\usepackage{mathtools}
\usepackage{dcolumn}
\usepackage{epsfig}
\usepackage{epstopdf}
\usepackage{bm}
\usepackage[left=1cm,right=1cm,top=2cm,bottom=2cm]{geometry}
\usepackage[usenames]{color}
\usepackage{colortbl}

\newcommand{\be}{\begin{equation}}
	\newcommand{\ee}{\end{equation}}

\newcommand{\bea}{\begin{eqnarray}}
	\newcommand{\eea}{\end{eqnarray}}
\newcommand{\beq}{\begin{equation}}
	\newcommand{\eeq}{\end{equation}}

\def\fun#1#2{\lower3.6pt\vbox{\baselineskip0pt\lineskip.9pt
		\ialign{$\mathsurround=0pt#1\hfil##\hfil$\crcr#2\crcr\sim\crcr}}}

\begin{document}
	\title{Deviation of nonradial geodesics in a static spherically symmetric space-time}
	\date{}
	\author{V.P. Vandeev$^*$, A.N. Semenova$^+$,}
	\maketitle
	\begin{center}
		{\it $^*$BALTIC STATE TECHNICAL UNIVERSITY «VOENMEH» named after D.F. Ustinov,\\ 1-Ya Krasnoarmeyskaya Ulitsa, 1, St Petersburg, 190005, Russia}
		
		{\it $^+$Petersburg Nuclear Physics Institute of National Research Centre ``Kurchatov Institute'',\\ Gatchina, 188300, Russia}
	\end{center}
\begin{abstract}
The article generalizes the description of tidal forces to the case of geodesics with non-zero angular momentum in the metric of static spherically symmetric black holes. We show that the
geodesic deviation equation can be diagonalized even with non-radial free motion of a test body in the gravitational field. We present expressions for the spatial components of the tidal force in a
spherically symmetric metric. We also solve geodesic deviation equation in the Schwarzschild metric and demonstrate how the presence of angular momentum and its magnitude affect the solution.
\end{abstract}

\section{\label{Int}Introduction}
The black hole properties are interesting for modern scientists, despite the fact that the first solution to describe a black hole was found in 1916 by Karl Schwarzschild \cite{MS},
which has a single parameter -- the mass of the black hole, they are still relevant both for theoretical and experimental research. Further, many solutions of Einstein's equations appeared,
which generalized Schwarzschild's solution: Reissner--Nordstr{\"{o}}m (electrically charged) black hole \cite{MR}, \cite{MN}, Kerr (rotating) black hole \cite{MK}, Kerr--Newman (charged and rotating)
black hole \cite{MKN}. Of particular interest are the properties of solutions to Einstein's equations in the presence of various matter. Examples are Kottler metric \cite{MKot}, which solves the GR
equations when the cosmological constant plays the role of matter and Kiselev metric, where matter is represented by the quintessence \cite{Kis}.

The main subject of interest for us is geodesic deviation. The first work devoted to solving geodesic deviation equation was \cite{SSSS}, there the behavior of the geodesic deviation
vector in the Schwarzschild metric was obtained. Despite the fact that modern studies of tidal forces are devoted to exotic solutions of the Einstein's equations: tidal effects in regular black holes
\cite{TFrbh}, tidal force in Schwarzschild spacetime in holographic massive gravity \cite{MG}, deviation vector behavior in 4D-Einstein--Gauss--Bonnet spacetime \cite{EGB}, in this paper we concentrate
on solving the problem of tidal forces in the metric of a spherically symmetric spacetime. However unlike articles \cite{TFrn}-\cite{TFkm} we consider non-radial geodesics.

Tidal effects play an important role in astrophysical phenomena. For example, tidal disruption event (or sometimes called tidal disruption flare) is that the star is destructed by the gravity of a supermassive black hole. Tidal forces acting on a star in the Schwarzschild gravitational field are studied in \cite{TDE}. The study of tidal effects in axisymmetric spactimes in the vicinity
of rotation axis, for example, in the Kerr metric \cite{KerrTF}, can make it possible to describe the formation of astrophysical jets. The description of relativistic flow formation through tidal acceleration
analysis is presented in \cite{J1}, \cite{J2}, \cite{J3}.

In our work the geodesic deviation equation depends on two integrals of motion: energy $E$ and angular momentum $L$. We demonstrate how geodesic deviation equation can be diogonalized
for static spherically symmetric metric defined by one function of radial variable. Such class of metrics include Schwarzschild, Reissner--Nordstr{\"{o}}m, Kottler, Kiselev spacetime etc. Geodesic
deviation equation with nonzero angular momentum have not previously been considered in works devoted to the study of tidal forces in static metrics. Probably the reason for this is the
impossibility of expressing the solutions of differential equations of geodesic deviation in elementary functions. We found that these equations satisfy Fuchs’ theorem, which allows us to construct
solutions in the vicinity of the points of interest using the Frobenius method. It is well known that in Schwarzschild spacetime a test body falling towards the event horizon of BH experiences stretching
along the radial direction and compression along the both angular directions \cite{GR}. We give a quantitative answer to the question of the angular momentum influence on tidal force components and on the
behavior of geodesic deviation equation solution. The effect of angular momentum on tidal forces is especially important for description of tidal destruction event, because the motion of stars
in the gravitational field of a black hole is non-radial. It should also be noted that the relative tidal acceleration between the two gravitational wave detectors was used to study the properties of
gravitons \cite{L1}, \cite{L2}.

The article is organized as follows. Seс.~\ref{GE} describes a system of geodesic equations that helps to define the velocity  4-vector of a freely falling test body. In Seс.~\ref{GDE} we consider the
geodesic deviation equation and describe the procedure for its diagonalization in a spherically symmetric spacetime. In Sec.~\ref{Tforce} we apply the expressions from the previous sections
to the Schwarzschild metric, then we solve geodesic deviation equation and compare the solutions for radial geodesics ($L=0$) with solutions for non-radial curves of free motion ($L\ne0$).
In the subsections of Sec.~\ref{Tforce} we present the numerical solutions of the geodesic deviation equation, local solutions in the vicinity of zero and infinity in the form of generalized power
series. Sec.~\ref{Conc} concludes our study, there we summarize and emphasize new results of our work and describe the range of
problems that can be solved within the framework of the tidal forces study following our work. We use the metric signature $(+,-,-,-)$ and set the speed of light $c$ and Newtonian gravitational constant
$G$ to $1$ throughout this paper.

\section{\label{GE}Geodesics in static spherically symmetric spacetime}
We consider spacetime with line element of static spherically symmetric black hole given by
\begin{equation}\label{metric}
ds^2=g_{\mu\nu}dx^{\mu}dx^{\nu}=f(r)dt^{2}-\frac{dr^2}{f(r)}-r^2\left(d\theta^2+\sin^2\theta d\phi^2\right).
\end{equation}
At this stage we do not specify the form of the function $f(r)$.
In this metric there are four meaningful geodesic equations \cite{mtbh} for all coordinates:
\begin{equation}\label{teq}
u^0=\frac{dt}{d\tau}=\frac{E}{f(r)},
\end{equation}
\begin{equation}\label{req}
\left(u^1\right)^2=\left(\frac{dr}{d\tau}\right)^2=E^2-f(r)\left(1+\frac{L^2}{r^2}\right),
\end{equation}
\begin{equation}\label{theq}
u^2=\frac{d\theta}{d\tau}=0,
\end{equation}
\begin{equation}\label{pheq}
u^3=\frac{d\phi}{d\tau}=\frac{L}{r^2},
\end{equation}
where $\tau$ is the proper time, $E$ and $L$ are energy and angular momentum of a freely moving test body respectively. And it should be noted that dynamics of the azimuthal variable is trivial due to the
spherical symmetry of spacetime, and the dynamics of the polar variable is nontrivial because there is angular momentum $L$. The set of expressions (\ref{teq})-(\ref{pheq}) forms a unit covariant
4-velocity vector $u^{\mu}$ tangent to the geodesic. Also it should be noted that work considers only timelike geodesics.

\section{\label{GDE}Geodesic deviation equation}
 Below we consider geodesic deviation equation. As it is well known \cite{GR}, the equation for the spacelike geodesic deviation vector $\tilde{\xi}^{\mu}$ is given by
\begin{equation}\label{deq}
\frac{D^2\tilde{\xi}^{\mu}}{d\tau^2}=R^{\mu}_{.\nu\alpha\beta}u^{\nu}u^{\alpha}\tilde{\xi}^{\beta},
\end{equation}
where $\frac{D^2}{d\tau^2}$ is covariant derivative along the geodesic, $R^{\mu}_{.\nu\alpha\beta}$ is Riemann curvature tensor and $u^{\nu}$ is the unit vector of 4-velocity tangent to the geodesic.
$\tilde{\xi}^{\mu}$ describes the distance between two infinitesimally close free moving particles.

The nonzero components of the Riemann tensor are calculated by the metric (\ref{metric}) and have the form
\begin{subequations}
\begin{equation}
R^{0}_{.101}=-\frac{f''}{2f},\\
\end{equation}
\begin{equation}
R^{0}_{.202}=R^{1}_{.212}=-\frac{f'r}{2},\\
\end{equation}
\begin{equation}
R^{0}_{.303}=R^{1}_{.313}=-\frac{f'r\sin^2{\theta}}{2},\\
\end{equation}
\begin{equation}
R^{2}_{.323}=\left(1-f\right)\sin^2{\theta}.
\end{equation}
\end{subequations}
It is seen that on the right-hand side of the Eq.~(\ref{deq}) there is a matrix $P^{\mu}_{.\beta}\equiv R^{\mu}_{.\nu\alpha\beta}u^{\nu}u^{\alpha}$ which is

\begin{equation}\label{matrix}
\begin{pmatrix}
\frac{\dot{r}^2f''}{2f}+\frac{\chi L^2 \sin^2\theta}{r^2}&-\frac{E\dot{r}f''}{2f^2}                          &0                                          &-\frac{\chi E L \sin^2\theta}{f}\\
\frac{E\dot{r}f''}{2}                                    &\frac{\chi L^2 \sin^2\theta}{r^2}-\frac{E^2f''}{2f}&0                                          &-\chi L \dot{r}\sin^2\theta\\
0                                                        &0                                                  &\frac{(f-1)L^2\sin^2\theta}{r^4}-\chi\omega&0\\
\frac{\chi E L}{r^2}                                     &-\frac{\chi L \dot{r}}{fr^2}                       &0                                          &-\chi\omega
\end{pmatrix},
\end{equation}

where
\begin{subequations}
\begin{equation}\label{r1}
\chi=\frac{f'}{2r},
\end{equation}
\begin{equation}\label{r2}
\omega=1+\frac{L^2}{r^2},
\end{equation}
\begin{equation}\label{r3}
\dot{r}=u^1=\frac{dr}{d\tau}=\sqrt{E^2-f\left(1+\frac{L^2}{r^2}\right)}.
\end{equation}
\end{subequations}
Matrix (\ref{matrix}) has spectrum
\begin{subequations}
\begin{equation}\label{evt}
\lambda_t=0,
\end{equation}
\begin{equation}\label{evr}
\lambda_r=\frac{\chi L^2\sin^2\theta}{r^2}-\frac{f''\omega}{2},
\end{equation}
\begin{equation}\label{evaz}
\lambda_\theta=\frac{(f-1)L^2\sin^2\theta}{r^4}-\chi\omega,
\end{equation}
\begin{equation}\label{evpol}
\lambda_\phi=\frac{\chi L^2\sin^2\theta}{r^2}-\chi\omega.
\end{equation}
\end{subequations}
Therefore, a tetrad basis for the free fall frame of reference can be constructed. It has form
\begin{subequations}
\begin{equation}\label{tett}
e_{t}^{\mu}=\frac{1}{\sqrt{1+\frac{L^2\cos^2\theta}{r^2}}}\left(\frac{E}{f},\dot{r},0,\frac{L}{r^2}\right),
\end{equation}
\begin{equation}\label{tetr}
e_{r}^{\mu}=\frac{1}{\sqrt{1+\frac{L^2}{r^2}}}\left(-\frac{\dot{r}}{f},-E,0,0\right),
\end{equation}
\begin{equation}\label{tetaz}
e_{\theta}^{\mu}=\left(0,0,\frac{1}{r},0\right),
\end{equation}
\begin{equation}\label{tetpol}
e_{\phi}^{\mu}=\frac{L\sin\theta}{r\sqrt{1+\frac{L^2\cos^2\theta}{r^2}}\sqrt{1+\frac{L^2}{r^2}}}\left(\frac{E}{f},\dot{r},0,\frac{1+\frac{L^2}{r^2}}{L\sin^2{\theta}}\right).\\
\end{equation}
\end{subequations}
This tetrad set $e^{\mu}_{\alpha}$ satisfy normalization condition ${e^{\mu}_{\alpha}\:e^{\nu}_{\beta}\:g_{\mu\nu}=\eta_{\alpha\beta}}$ with Minkowski metric $\eta_{\alpha\beta}=\text{diag}\left(1,-1,-1,-1\right)$.
The geodesic deviation vector $\tilde{\xi}^{\mu}$ can be substituted as
\begin{equation}\label{lt}
    \tilde{\xi}^{\mu}=e^{\mu}_{\nu}\:\xi^{\nu}.
\end{equation}
Thus, the meaningful components of Eq. (\ref{deq}) in the new coordinate system are
\begin{equation}\label{radeq}
\ddot{\xi}^r=\left[-\frac{f''}{2}\left(1+\frac{L^2}{r^2}\right)+\frac{f'}{2r}\frac{L^2}{r^2}\right]\xi^r,
\end{equation}
\begin{equation}\label{azeq}
\ddot{\xi}^\theta=\left[-\frac{f'}{2r}\left(1+\frac{L^2}{r^2}\right)+\frac{f-1}{r^2}\frac{L^2}{r^2}\right]\xi^\theta,
\end{equation}
\begin{equation}\label{poleq}
\ddot{\xi}^\phi=-\frac{f'}{2r}\xi^\phi.
\end{equation}
These equations are the diagonal form of Eq.~(\ref{deq}). It is worth noting that the dependence on the $\theta$ disappeared from Eqs.~(\ref{radeq})--(\ref{poleq}) because there is no dynamics along the
direction of the azimuthal angle $\theta$ according to Eq.~(\ref{theq}) and geodesics lie in the equatorial plane $\theta=\frac{\pi}{2}$.

The diagonal form of Eq.~(\ref{deq}) for a static spherically symmetric spacetime for non-radial geodesics was obtained in \cite{SSSS} but a slightly more complex mathematical apparatus was used there,
although in fact it is sufficient to use the methods of linear algebra. For $L=0$ Eqs.~(\ref{radeq})--(\ref{poleq}) coincide with Eqs.~(14) and (15) of Ref. \cite{TFrn}.

To illustrate the result obtained, we can use superposition of charged black hole and (Anti)-de Sitter spacetime, its metric is
\begin{equation}\label{srnk}
ds^2=f(r)dt^{2}-\frac{dr^2}{f(r)}-r^2\left(d\theta^2+\sin^2\theta d\phi^2\right),
\end{equation}
where
\begin{equation}\label{fsrnk}
f(r)=1-\frac{2M}{r}+\frac{q^2}{r^2}-\frac{\Lambda}{3}r^2.
\end{equation}
Using (\ref{fsrnk}) in Eqs.~(\ref{radeq})--(\ref{poleq}) we get
\begin{subequations}\label{ill}
\begin{equation}\label{illr}
\ddot{\xi}^r=\left[\frac{2M}{r^3}-\frac{3q^2}{r^4}+\frac{\Lambda}{3}+\frac{L^2}{r^2}\left(\frac{3M}{r^3}-\frac{4q^2}{r^4}\right)\right]\xi^r,
\end{equation}
\begin{equation}\label{illaz}
\ddot{\xi}^\theta=\left[-\frac{M}{r^3}+\frac{q^2}{r^4}+\frac{\Lambda}{3}+\frac{L^2}{r^2}\left(-\frac{3M}{r^3}+\frac{2q^2}{r^4}\right)\right]\xi^\theta,
\end{equation}
\begin{equation}\label{illpol}
\ddot{\xi}^\phi=\left[-\frac{M}{r^3}+\frac{q^2}{r^4}+\frac{\Lambda}{3}\right]\xi^\phi.
\end{equation}
\end{subequations}
\section{\label{Tforce}Tidal force}
\begin{figure}[h!]
\center{\includegraphics[width = 15 cm]{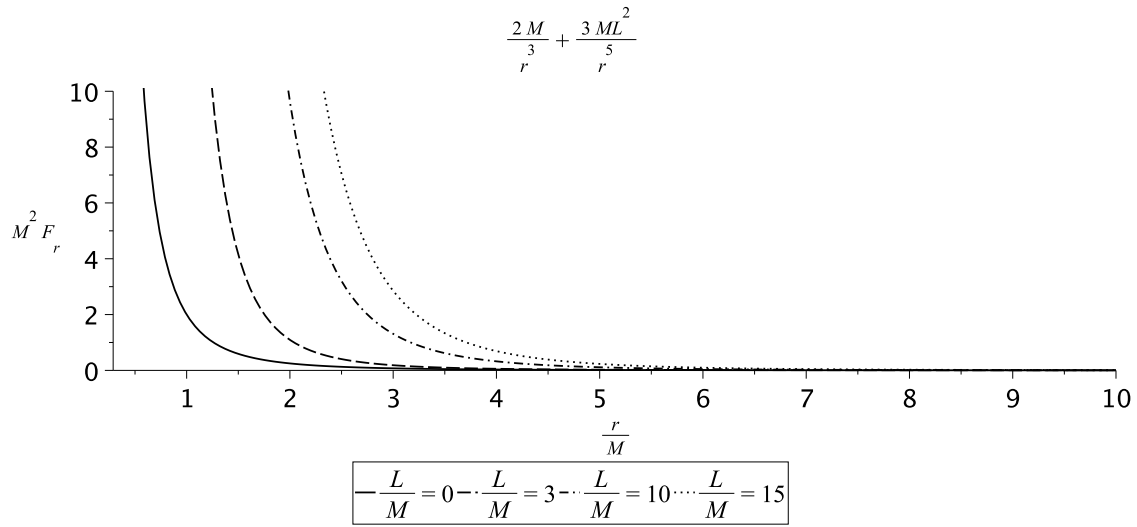}
\caption{\footnotesize{Radial tidal force component for Schwarzschild BH with different choices of $L$.}}\label{Fr}}\end{figure}

\begin{figure}[h!]
\center{\includegraphics[width = 15 cm]{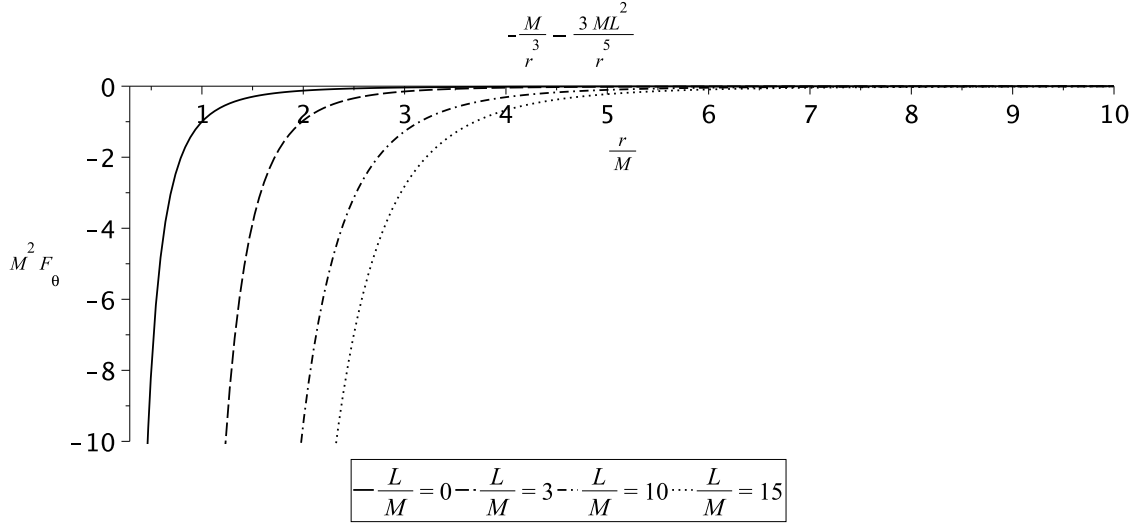}
\caption{\footnotesize{Azimuthal tidal force component for Schwarzschild BH with different choices of $L$.}}\label{Faz}}
\end{figure}

Expressions (\ref{ill}) are rather complicated due to the large number of parameters, therefore, below we restrict ourselves to an electrically neutral black hole with no cosmological constant
($q=0$ and $\Lambda=0$). In other words, we consider the Schwarzschild metric
\begin{equation}\label{prr}
\ddot{\xi}^r=\left[\frac{2M}{r^3}+\frac{3ML^2}{r^5}\right]\xi^r,
\end{equation}
\begin{equation}\label{praz}
\ddot{\xi}^\theta=\left[-\frac{M}{r^3}-\frac{3ML^2}{r^5}\right]\xi^\theta,
\end{equation}
\begin{equation}\label{prpol}
\ddot{\xi}^\phi=-\frac{M}{r^3}\xi^\phi.
\end{equation}
This case for $L=0$ was considered in \cite{GR}. Note that the presence of angular momentum makes the expressions for the angular components (\ref{praz}) and (\ref{prpol}) of the tidal force different.
The presence of terms with angular momentum in the radial (\ref{prr}) and azimuthal (\ref{praz}) components of the tidal force does not change the monotonicity of these components depending on the radius $r$.
This can be seen in Figs.~\ref{Fr} and \ref{Faz}. Also it is obvious that all components of the tidal force are constant sign and do not vanish for finite $r$.

There is no need to depict the polar component of the tidal force (\ref{prpol}) because it is identical to the azimuthal one at $L=0$. It looks the same as the black solid curve in Fig.~\ref{Faz}. Also it
should be noted that in the case of Schwarzschild metric we do not consider very large values of the angular momentum $L$ because Eq.~(\ref{req}) for $L\to \infty$ has form of a power series
\begin{equation}
\frac{dr}{d\tau}=\frac{L}{r}\sqrt{\frac{2M-r}{r}}+O\left(L^{-1}\right),
\end{equation}
which is defined only under the black hole horizon for $r\le 2M$. The absence of geodesics above the horizon $r>2M$ naturally leads to the absence of their deviation.
\subsection{\label{sol}Solutions of the geodesic deviation equations}
Using Eq.~(\ref{req}) we pass from differentiation with respect to proper time $\tau$ to differentiation with respect to dimensionless radial variable $\rho=\frac{r}{M}$ in Eqs.~(\ref{prr})--(\ref{prpol})
\begin{equation}\label{eqr}
{\xi^r}''+Q(\rho){\xi^r}'+P_r(\rho)\xi^r=0,
\end{equation}
\begin{equation}\label{eqaz}
{\xi^\theta}''+Q(\rho){\xi^\theta}'+P_\theta(\rho)\xi^\theta=0,
\end{equation}
\begin{equation}\label{eqpol}
{\xi^\phi}''+Q(\rho){\xi^\phi}'+P_\phi(\rho)\xi^\phi=0.
\end{equation}
Where $l=\frac{L}{M}$ is dimensionless angular momentum and the following notation is used for abbreviation
\begin{subequations}\label{func}
\begin{equation}
Q(\rho)=-\frac{\rho^2-l^2\rho+3l^2}{(E^2-1)\rho^4+2\rho^3-l^2\rho^2+2l^2\rho},
\end{equation}
\begin{equation}
P_r(\rho)=-\frac{2\rho^2+3l^2}{(E^2-1)\rho^5+2\rho^4-l^2\rho^3+2l^2\rho^2},
\end{equation}
\begin{equation}
P_\theta(\rho)=\frac{\rho^2+3l^2}{(E^2-1)\rho^5+2\rho^4-l^2\rho^3+2l^2\rho^2},
\end{equation}
\begin{equation}
P_\phi(\rho)=\frac{1}{(E^2-1)\rho^3+2\rho^2-l^2\rho+2l^2}.
\end{equation}
\end{subequations}
It should be noted that for $l=0$ the solutions of the Eqs.~(\ref{eqr})--(\ref{eqpol}) are expressed in terms of elementary functions
\begin{equation}\label{elsolr}
\xi^r=A\alpha(\rho)+\frac{B}{\gamma^4}\left[6+\gamma^2\rho+\frac{3\alpha(\rho)}{\gamma}\ln\left(\frac{\alpha(\rho)-\gamma}{\alpha(\rho)+\gamma}\right)\right],
\end{equation}
\begin{equation}\label{elsola}
\xi^\theta=\xi^\phi=\rho\bigg(C+D\alpha(\rho)\bigg),
\end{equation}
where
\begin{equation}
\gamma=\sqrt{E^2-1},\;\alpha(\rho)=\sqrt{E^2-1+\frac{2}{\rho}},
\end{equation}
and $A, B, C$ and $D$ are integration constants. These solutions were firstly obtained in Ref. \cite{SSSS}.

These solutions in the vicinity of $\rho=0$ are
\begin{subequations}\label{zerorsing}
\begin{equation}\label{zerorsingr}
\xi^{r}(\rho)=\frac{A\sqrt{2}}{\sqrt{\rho}}+O\left(\sqrt{\rho}\right)\;\text{for}\;\rho\to0,
\end{equation}
\begin{equation}\label{zerorsingaz}
\xi^{\theta}(\rho)=\sqrt{2}D_\theta\sqrt{\rho}+O\left(\rho\right)\;\text{for}\;\rho\to0,
\end{equation}
\begin{equation}\label{zerorsingpol}
\xi^{\phi}(\rho)=\sqrt{2}D_\phi\sqrt{\rho}+O\left(\rho\right)\;\text{for}\;\rho\to0.
\end{equation}
\end{subequations}

The same solutions in the vicinity of $\rho=\infty$ behave
\begin{subequations}\label{largersing}
\begin{equation}\label{largersingr}
\xi^{r}(\rho)=\frac{B}{E^2-1}\rho+O\left(1\right)\;\text{for}\;\rho\to\infty,
\end{equation}
\begin{equation}\label{largersingaz}
\xi^{\theta}(\rho)=\left(C_\theta+D_\theta\sqrt{E^2-1}\right)\rho+O\left(1\right)\;\text{for}\;\rho\to\infty,
\end{equation}
\begin{equation}\label{largersingpol}
\xi^{\phi}(\rho)=\left(C_\phi+D_\phi\sqrt{E^2-1}\right)\rho+O\left(1\right)\;\text{for}\;\rho\to\infty.
\end{equation}
\end{subequations}
Solutions of (\ref{eqr})--(\ref{eqpol}) can not be expressed in terms of elementary functions for the case $l\ne0$ so we will find them in the form of generalized series in the vicinity of points
$\rho=0$ and $\rho=\infty$. We will present a comparison of such solutions for non-zero angular momentum and Eqs.~(\ref{zerorsing}) and (\ref{largersing}) in the following sections.

\subsection{\label{numsol}Numerical solutions}
\begin{figure}[h!]
\center{\includegraphics[width = 15 cm]{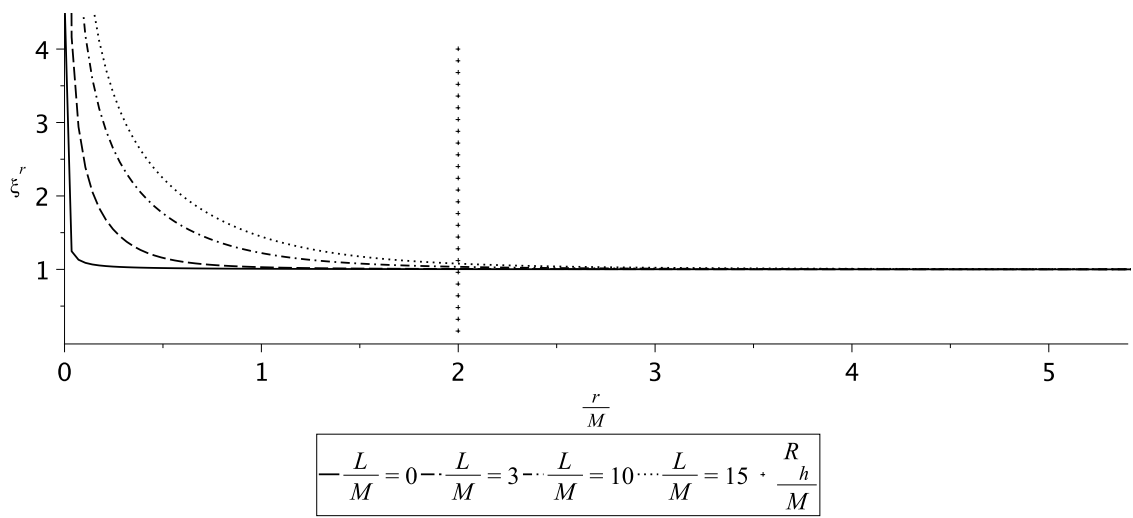}
\caption{\footnotesize{Dependence of $\xi^r$ on $\rho$ at $E=10$. Initial data are $\xi^{r}(10)=1$ and ${\xi^{r}}'(10)=0$.}}
\label{rfig}}
\end{figure}

Nevertheless, the solutions of Eqs.~(\ref{eqr})--(\ref{eqpol}) can be found numerically. In Fig.~\ref{rfig} we construct a numerical solution of the radial Eq.~(\ref{eqr}) for various values of the angular
momentum $l$. It can be seen that the black solid curve with $l=0$ in the vicinity of $\rho=0$ corresponds to (\ref{zerorsingr}).

\begin{figure}[h!]
\center{\includegraphics[width = 15 cm]{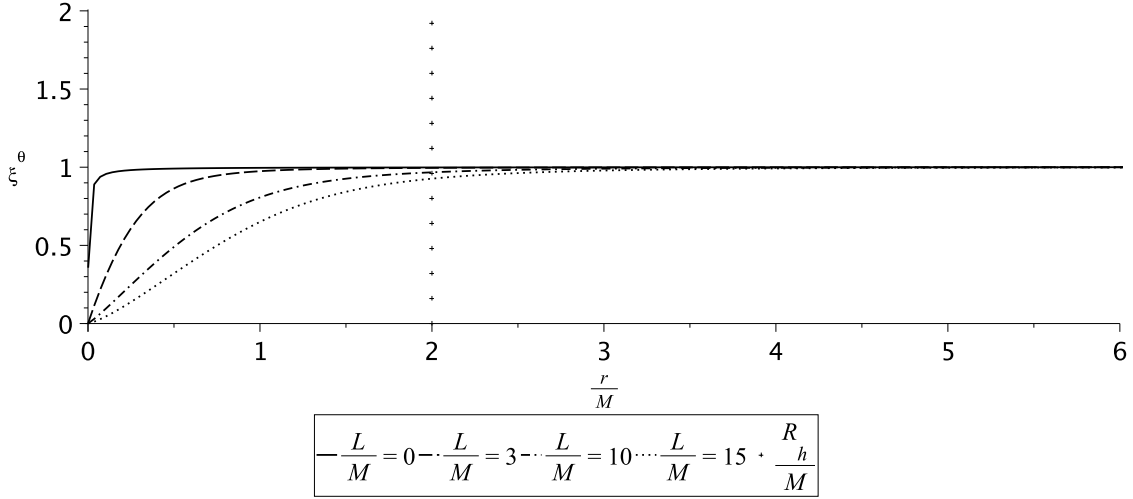}
\caption{\footnotesize{Dependence of $\xi^{\theta}$ on $\rho$ at $E=10$. Initial data are $\xi^{\theta}(10)=1$ and ${\xi^{\theta}}'(10)=0$.}}
\label{azfig}}
\end{figure}

In Fig.~\ref{azfig} we construct a numerical solution of the azimuthal Eq.~(\ref{eqaz}) for various values of the angular momentum $l$. It can be seen that the black solid curve with $l=0$ in the vicinity of
$\rho=0$ corresponds to (\ref{zerorsingaz}).

\begin{figure}[h!]
\center{\includegraphics[width = 15 cm]{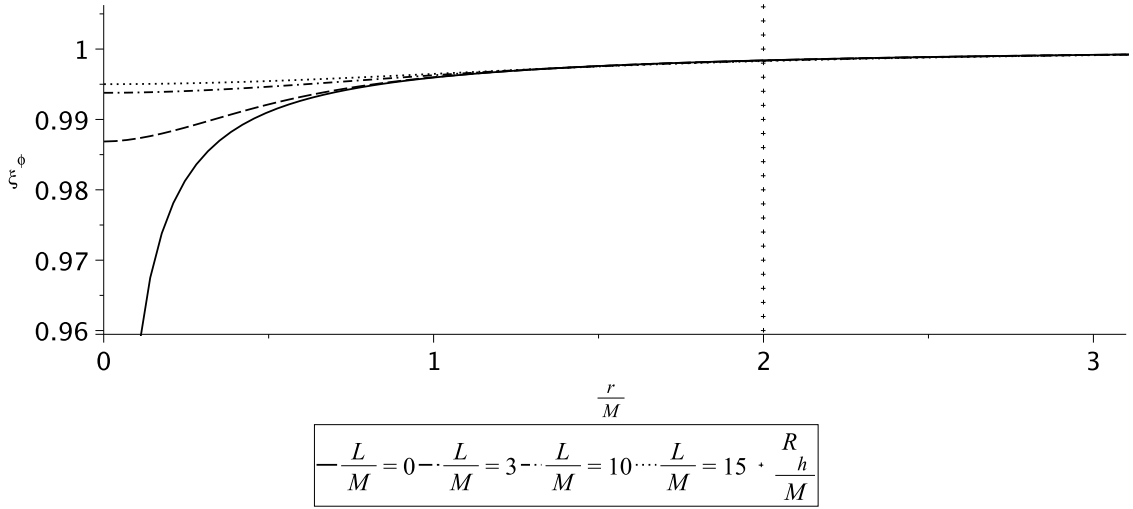}
\caption{\footnotesize{Dependence of $\xi^{\phi}$ on $\rho$ at $E=10$. Initial data are $\xi^{\phi}(10)=1$ and ${\xi^{\phi}}'(10)=0$.}}
\label{polfig}}
\end{figure}

In Fig.~\ref{polfig} we construct a numerical solution of the polar Eq.~(\ref{eqpol}) for various values of the angular momentum $l$. It can be seen that the black solid curve with $l=0$ in the vicinity of
$\rho=0$ corresponds to (\ref{zerorsingpol}).

\subsection{\label{locsol}Local behavior of solutions}
Below we solve Eqs.~(\ref{eqr})--(\ref{eqpol}) in the vicinity of points $\rho=0$, $\rho=\infty$. Since these equations cannot be solved in terms of elementary functions, we will construct solutions in
the form of power series.
\subsubsection{\label{secsubsing}In the BH singularity vicinity}
Here we represent functions (\ref{func}) in the form of a power series near $\rho=0$
\begin{subequations}\label{funczero}
\begin{equation}\label{qsing}
Q(\rho)=-\frac{3}{2\rho}-\frac{1}{4}+\frac{8-l^2}{8l^2}\rho+\frac{12E^2-l^2}{16l^2}\rho^2+O\left(\rho^3\right),
\end{equation}
\begin{equation}\label{prsing}
P_r(\rho)=-\frac{3}{2\rho^2}-\frac{3}{4\rho}+\frac{4-3l^2}{8l^2}+\frac{12E^2-3l^2+4}{16l^2}\rho+O\left(\rho^2\right),
\end{equation}
\begin{equation}\label{pazsing}
P_\theta(\rho)=\frac{3}{2\rho^2}+\frac{3}{4\rho}-\frac{8-3l^2}{8l^2}-\frac{12E^2-3l^2+8}{16l^2}\rho+O\left(\rho^2\right),
\end{equation}
\begin{equation}\label{ppolsing}
P_\phi(\rho)=\frac{1}{2l^2}+\frac{\rho}{4l^2}+\frac{l^2-4}{8l^4}\rho^2-\frac{4E^2-l^2+4}{16l^4}\rho^3+O\left(\rho^4\right).
\end{equation}
\end{subequations}
It follows from the form of these series that the Eqs.~(\ref{eqr})--(\ref{eqpol}) satisfy Fuchs' theorem. This means that their solutions can be found using generalized power series by the Frobenius method.
Therefore, each of them has two linearly independent solutions
\begin{subequations}\label{subs1}
\begin{equation}
\xi^j_1=\sum_{n=0}^{\infty}c^{j}_n\rho^{n+\zeta_1},
\end{equation}
\begin{equation}
\xi^j_2=\sum_{n=0}^{\infty}d^{j}_n\rho^{n+\zeta_2},
\end{equation}
\end{subequations}
where $j=r,\theta,\phi$. And $\zeta_1,\zeta_2$ are roots of the quadratic equation $\zeta(\zeta-1)+a^j_{-1}\zeta+b^j_{-2}=0$ (where $a^j_{-1}$ is coefficient of $z^{-1}$ in series (\ref{qsing})
and $b^j_{-2}$ is coefficient of $z^{-2}$ in series (\ref{prsing}), (\ref{pazsing}) or (\ref{ppolsing}) correspondingly). Consider each of the Eqs.~(\ref{eqr})--(\ref{eqpol}) separately.
\begin{enumerate}
\item For radial Eq.~(\ref{eqr}) there is a quadratic equation for powers $\zeta(\zeta-1)-\frac{3}{2}\zeta-\frac{3}{2}=0$. They are $\zeta_1=3,\zeta_2=-\frac{1}{2}$. This allows us to substitute the series
(\ref{subs1}) with the leading powers into Eq.~(\ref{eqr}), equate the coefficients at all powers of $\rho$ to zero. We have expressed all unknown coefficients in terms of $c^r_0$ and $d^r_0$, which are
assumed to be equal to one, therefore the solutions to the Eq.~(\ref{eqr}) are
\begin{subequations}\label{soleq22}
\begin{equation}
\xi^r_1(\rho)=\rho^3+\frac{\rho^4}{3}+\frac{\left(8l^2-21\right)\rho^5}{66l^2}-\frac{\left(132E^2-40l^2+105\right)\rho^6}{858l^2}+O\left(\rho^7\right).
\end{equation}
\begin{equation}
\xi^r_2(\rho)=\frac{1}{\sqrt{\rho}}-\frac{\rho^{\frac{1}{2}}}{4}-\frac{\rho^{\frac{3}{2}}}{32}+\frac{\left(32E^2-l^2\right)\rho^{\frac{5}{2}}}{128l^2}+O\left(\rho^{\frac{7}{2}}\right),
\end{equation}
\end{subequations}
And the general solution to Eq. (\ref{eqr}) is a linear combination
\begin{equation}\label{gensoleq22}
\xi^r(\rho)=A_r\xi^r_1(\rho)+B_r\xi^r_2(\rho),
\end{equation}
where $A_r$ and $B_r$ are arbitrary constants.
\item For azimuthal Eq.~(\ref{eqaz}) there is a quadratic equation for powers $\zeta(\zeta-1)-\frac{3}{2}\zeta+\frac{3}{2}=0$. They are easy to find $\zeta_1=\frac{3}{2},\zeta_2=1$. This allows us to
substitute the series (\ref{subs1}) with the leading powers into Eq.~(\ref{eqaz}), equate the coefficients at all powers of $\rho$ to zero. We have expressed all unknown coefficients in terms
of $c^\theta_0$ and $d^\theta_0$, which are assumed to be equal to one, therefore the solutions to the Eq.~(\ref{eqaz}) are
 \begin{subequations}\label{soleq23}
\begin{equation}
\xi^\theta_1(\rho)=\rho^{\frac{3}{2}}-\frac{\rho^{\frac{5}{2}}}{4}-\frac{\left(5l^2+16\right)\rho^{\frac{7}{2}}}{160l^2}-\frac{\left(160E^2+35l^2-368\right)\rho^{\frac{9}{2}}}{4480l^2}+\left(\rho^{\frac{11}{2}}\right),
\end{equation}
\begin{equation}
\xi^\theta_2(\rho)=\rho-\rho^2+\frac{\rho^4}{5l^2}+\frac{\left(5E^2-1\right)\rho^5}{70l^2}+O\left(\rho^6\right).
\end{equation}
\end{subequations}
And the general solution to Eq. (\ref{eqaz}) is a linear combination
\begin{equation}\label{gensoleq23}
\xi^\theta(\rho)=A_\theta\xi^\theta_1(\rho)+B_\theta\xi^\theta_2(\rho),
\end{equation}
where $A_\theta$ and $B_\theta$ are arbitrary constants.
\item For polar Eq.~(\ref{eqpol}) there is a quadratic equation for powers $\zeta(\zeta-1)-\frac{3}{2}\zeta=0$. They are $\zeta_1=\frac{5}{2},\zeta_2=0$. This allows us to substitute the series
(\ref{subs1}) with the leading powers into Eq.~(\ref{eqpol}), equate the coefficients at all powers of $\rho$ to zero. We have expressed all unknown coefficients in terms of $c^\phi_0$ and $d^\phi_0$,
which are assumed to be equal to one, therefore the solutions to the Eq.~(\ref{eqpol}) are
\begin{subequations}\label{soleq24}
\begin{equation}
\xi^\phi_1(\rho)=\rho^{\frac{5}{2}}+\frac{5\rho^{\frac{7}{2}}}{28}+\frac{\left(5l^2-32\right)\rho^{\frac{9}{2}}}{96l^2}-\frac{5\left(224E^2-35l^2+160\right)\rho^{\frac{11}{2}}}{9856l^2}+O\left(\rho^{\frac{11}{2}}\right),
\end{equation}
\begin{equation}
\xi^\phi_2(\rho)=1+\frac{\rho^2}{2l^2}-\frac{\rho^4}{8l^4}-\frac{E^2\rho^5}{25l^4}+\left(\rho^6\right).
\end{equation}
\end{subequations}
And the general solution to Eq.~(\ref{eqpol}) is a linear combination
\begin{equation}\label{gensoleq24}
\xi^\theta(\rho)=A_\phi\xi^\phi_1(\rho)+B_\phi\xi^\phi_2(\rho),
\end{equation}
where $A_\phi$ and $B_\phi$ are arbitrary constants.
\end{enumerate}
Leaving only the leading members in (\ref{gensoleq22}), (\ref{gensoleq23}) and (\ref{gensoleq24}) we have
\begin{subequations}\label{lmsol}
\begin{equation}\label{lmsolr}
\xi^{r}(\rho)=\frac{B_r}{\sqrt{\rho}}+O\left(\rho^{\frac{1}{2}}\right)\;\text{for}\;\rho\to0,
\end{equation}
\begin{equation}\label{lmsolaz}
\xi^{\theta}(\rho)=B_\theta\rho+O\left(\rho^{\frac{3}{2}}\right)\;\text{for}\;\rho\to0,
\end{equation}
\begin{equation}\label{lmsolpol}
\xi^{\phi}(\rho)=B_\phi+O\left(\rho^2\right)\;\text{for}\;\rho\to0.
\end{equation}
\end{subequations}
Comparing expressions (\ref{zerorsing}) and (\ref{lmsol}) it is seen that angular momentum presence does not affect the behavior of the radial geodesic deviation vector component in the vicinity of zero.
At small $\rho$ (\ref{zerorsingr}) and (\ref{lmsolr}) behave in the leading order as inverse square root $\xi^{r}=\frac{C}{\sqrt{\rho}}$. So near physical singularity the test free-falling body experiences tidal stretching
along the radial direction regardless of angular momentum presence. Note that Eq.~(\ref{lmsolr}) corresponds to numerical solutions in Fig.~\ref{rfig} for $l=3,10,15$.

However, the behavior of the angular components changed with the appearance of the angular momentum $l$. The azimuthal component (\ref{zerorsingaz}) of geodesic deviation vector without $l$ behaves
as $\xi^{\theta}=C\sqrt{\rho}$, but with $l\ne0$ it behaves linearly $\xi^{\theta}=C\rho$ (\ref{lmsolaz}) at small $\rho$. Note that Eq.~(\ref{lmsolaz}) corresponds to numerical solutions in Fig.~\ref{azfig}
for $l=3,10,15$.

Also the polar component of geodesic deviation vector changes its dependence on the radial variable $\rho$ when angular momentum appears. Without angular momentum the $\phi$-component behaves like
$\theta$-component (\ref{zerorsingpol}) as $\xi^{\phi}=C\sqrt{\rho}$. But when there is angular momentum, the polar component near $\rho=0$ becomes constant (\ref{lmsolpol}). Note that Eq.~(\ref{lmsolpol})
corresponds to numerical solutions in Fig.~\ref{polfig} for $l=3,10,15$.

\subsubsection{\label{secsubinf}Infinitely far from BH}
To determine the asymptotic behavior of equations solutions  (\ref{eqr})--(\ref{eqpol}) at large $\rho$ we need to replace the variable $\rho=\frac{1}{s}$ and consider the solution in the
vicinity of the point $s=0$. New form of Eqs. (\ref{eqr})--(\ref{eqpol}) is
\begin{equation}\label{eqrinf}
{\xi^r}''+W(s){\xi^r}'+G_r(s)\xi^r=0,
\end{equation}
\begin{equation}\label{eqazinf}
{\xi^\theta}''+W(s){\xi^\theta}'+G_\theta(s)\xi^\theta=0,
\end{equation}
\begin{equation}\label{eqpolinf}
{\xi^\phi}''+W(s){\xi^\phi}'+G_\phi(s)\xi^\phi=0,
\end{equation}
where
\begin{subequations}\label{funcinf}
\begin{equation}
W(s)=\frac{2}{s}+\frac{1-l^2s+3l^2s^2}{(E^2-1)+2s-l^2s^2+2l^2s^3},
\end{equation}
\begin{equation}\label{Gr}
G_r(s)=-\frac{2+3l^2s^2}{(E^2-1)s+2s^2-l^2s^3+2l^2s^4},
\end{equation}
\begin{equation}\label{Gaz}
G_\theta(s)=\frac{1+3l^2s^2}{(E^2-1)s+2s^2-l^2s^3+2l^2s^4},
\end{equation}
\begin{equation}\label{Gpol}
G_\phi(s)=\frac{1}{(E^2-1)s+2s^2-l^2s^3+2l^2s^4}.
\end{equation}
\end{subequations}
Now we solve these equations in the vicinity of the point $s=0$, which corresponds to $\rho=\infty$. Therefore, we expand the functions (\ref{funcinf}) in power series for $s\to0$. These series are
\begin{subequations}\label{funcinfs}
\begin{equation}\label{Wser}
W(s)=\frac{2}{s}+\frac{1}{E^2-1}-\left[\frac{l^2(E^2-1)+2}{(E^2-1)^2}\right]s+O(s^2),
\end{equation}
\begin{equation}\label{Grser}
G_r(s)=-\frac{2}{(E^2-1)s}+\frac{4}{(E^2-1)^2}-\left[\frac{3l^2(E^2-1)^2+2l^2(E^2-1)+8}{(E^2-1)^3}\right]s+O\left(s^2\right),
\end{equation}
\begin{equation}\label{Gazser}
G_\theta(s)=\frac{1}{(E^2-1)s}-\frac{2}{(E^2-1)^2}+\left[\frac{3l^2(E^2-1)^2+l^2(E^2-1)+4}{(E^2-1)^3}\right]s+O\left(s^2\right),
\end{equation}
\begin{equation}\label{Gpolser}
G_\phi(s)=\frac{1}{(E^2-1)s}-\frac{2}{(E^2-1)^2}+\left[\frac{l^2(E^2-1)+4}{(E^2-1)^3}\right]s+O\left(s^2\right).
\end{equation}
\end{subequations}

According to the Fuchs' theorem Frobenius method is applicable again and the point $s=0$ is a regular singular point for Eqs.~(\ref{eqrinf})--(\ref{eqpolinf}). For all series (\ref{Grser})--(\ref{Gpolser})
coefficient ${b_{-2}=0}$ since the functions (\ref{Gr})--(\ref{Gpol}) in $s=0$ have first order pole and coefficient $a_{-1}=2$ of (\ref{Wser}). Thus, the quadratic equation for the leading powers
has form $\zeta(\zeta-1)+2\zeta=0$. Leading powers are $\zeta_1=0$ and $\zeta_2=-1$, their difference is equal to an integer. Therefore, two linearly independent solutions have the form
\begin{subequations}\label{subs2}
\begin{equation}\label{subs2a}
\xi^j_1(s)=\sum_{n=0}^{\infty}c^{j}_ns^{n+\zeta_1},
\end{equation}
\begin{equation}\label{subs2b}
\xi^j_2(s)=\sum_{n=0}^{\infty}d^{j}_ns^{n+\zeta_2}+A\xi^j_1(s)\ln(s),
\end{equation}
\end{subequations}
where $j=r,\theta,\phi$ corresponds to different equations.
To find the coefficients $\{c^j_n,d^j_n\}_{n=1}^{\infty}$ and $A$, one need to substitute these series into the Eqs.~(\ref{eqrinf})--(\ref{eqpolinf}) and after simplifying like terms
equate the coefficients at all powers of $s$ to zero.
\begin{enumerate}
\item By substituting expression (\ref{subs2a}) in Eq.~(\ref{eqrinf}), the coefficients $c^r_j,\;j\ge1$  can be expressed in terms of $c^r_0$, equating which to one we will find the first solution
\begin{equation}\label{soleq40a}
\xi^r_1(s)=1+\frac{s}{E^2-1}-\frac{s^2}{2\left(E^2-1\right)^2}+\frac{\left(E^4l^2-E^2l^2+2\right)s^3}{4\left(E^2-1\right)^3}+O\left(s^4\right).
\end{equation}
To find the second linearly independent solution, substitute expression (\ref{subs2b}) into Eq.~(\ref{eqrinf}), assuming $\xi^r_1(s)$ as a solution.
Equating the coefficients at all powers of $s$ to zero and setting $d^r_0=1$ and $d^r_1=0$ since $d^r_0$ is free coefficient, and $d^r_1$ is a factor at $s^{\zeta_1}$ in second solution, we get
\begin{multline}\label{soleq40b}
\xi^r_2(s)=\ln(s)\left[\frac{3}{E^2-1}+\frac{3s}{\left(E^2-1\right)^2}-\frac{3s^2}{2(E^2-1)^3}\right]+\frac{1}{s}-\\-\frac{\left(E^2l^2-l^2+18\right)s}{2\left(E^2-1\right)^2}+\frac{\left(4E^4l^2-3E^2l^2-l^2+9\right)s^2}{4\left(E^2-1\right)^3}+O\left(s^3\right).
\end{multline}
Therefore, the general solution to Eq.~(\ref{eqrinf}) is a linear combination of (\ref{soleq40a}) and (\ref{soleq40b})
\begin{equation}\label{rsolinf}
\xi^r(s)=A_r\xi^r_1(s)+B_r\xi^r_2(s),
\end{equation}
where $A_r$ and $B_r$ are arbitrary constants.
\item By substituting expression (\ref{subs2a}) in Eq.~(\ref{eqazinf}), the coefficients $c^\theta_j,\;j\ge1$  can be expressed in terms of $c^\theta_0$, equating which to one we will find the first
solution
\begin{equation}\label{soleq41a}
\xi^\theta_1(s)=1-\frac{s}{2\left(E^2-1\right)}+\frac{s^2}{2\left(E^2-1\right)^2}-\frac{\left(2E^4l^2-3E^2l^2+l^2+5\right)s^3}{8\left(E^2-1\right)^3}+O\left(s^4\right).
\end{equation}
To find the second linearly independent solution, substitute expression (\ref{subs2b}) into Eq.~(\ref{eqazinf}), assuming $\xi^\theta_1(s)$ as a solution.
Equating the coefficients at all powers of $s$ to zero and setting $d^\theta_0=1$ and $d^\theta_1=0$ since $d^\theta_0$ is free coefficient, and $d^\theta_1$ is a factor at $s^{\zeta_1}$ in
second solution, we get
\begin{equation}\label{soleq41b}
\xi^\theta_2(s)=\frac{1}{s}-\frac{l^2s}{2\left(E^2-1\right)}+\frac{l^2s^2}{2\left(E^2-1\right)^2}-\frac{\left(E^2l^4-l^4+5l^2\right)s^3}{8\left(E^2-1\right)^3}+O\left(s^4\right).
\end{equation}
Therefore, the general solution to Eq.~(\ref{eqazinf}) is a linear combination of (\ref{soleq41a}) and (\ref{soleq41b})
\begin{equation}\label{thetasolinf}
\xi^r(s)=A_\theta\xi^\theta_1(s)+B_\theta\xi^\theta_2(s),
\end{equation}
where $A_\theta$ and $B_\theta$ are arbitrary constants.
\item By substituting expression (\ref{subs2a}) in Eq.~(\ref{eqpolinf}), the coefficients $c^\phi_j,\;j\ge1$  can be expressed in terms of $c^\phi_0$, equating which to one we will find the first solution
\begin{equation}\label{soleq42a}
\xi^\phi_1(s)=1-\frac{s}{2\left(E^2-1\right)}+\frac{s^2}{2\left(E^2-1\right)^2}-\frac{\left(E^2l^2-l^2+5\right)s^3}{8\left(E^2-1\right)^3}+O\left(s^4\right).
\end{equation}
To find the second linearly independent solution, substitute expression (\ref{subs2b}) into Eq.~(\ref{eqpolinf}), assuming $\xi^\phi_1(s)$ as a solution.
Equating the coefficients at all powers of $s$ to zero and setting $d^\phi_0=1$ and $d^\phi_1=0$ since $d^\phi_0$ is free coefficient, and $d^r_1$ is a factor at $s^{\zeta_1}$ in second solution, we get
\begin{equation}\label{soleq42b}
\xi^\phi_2(s)=\frac{1}{s}-\frac{l^2s}{2\left(E^2-1\right)}+\frac{E^2l^2s^2}{2\left(E^2-1\right)^2}-\frac{\left(E^2l^4+5E^2l^2-l^4\right)s^3}{8\left(E^2-1\right)^3}+O\left(s^4\right).
\end{equation}
Therefore, the general solution to Eq.~(\ref{eqpolinf}) is a linear combination of (\ref{soleq42a}) and (\ref{soleq42b})
\begin{equation}\label{phisolinf}
\xi^r(s)=A_\phi\xi^\phi_1(s)+B_\phi\xi^\phi_2(s),
\end{equation}
where $A_\phi$ and $B_\phi$ are arbitrary constants.
\end{enumerate}

Leaving only the leading members in (\ref{rsolinf}), (\ref{thetasolinf}) and (\ref{phisolinf}) we have
\begin{subequations}
\begin{equation}
\xi^{r}(s)=B_r\left(\frac{1}{s}+\frac{3\ln(s)}{E^2-1}\right)+A_r+O\left(s\right),
\end{equation}
\begin{equation}
\xi^{\theta}(s)=\frac{B_\theta}{s}+A_\theta+O\left(s\right),
\end{equation}
\begin{equation}
\xi^{\phi}(s)=\frac{B_\phi}{s}+A_\phi+O\left(s\right).
\end{equation}
\end{subequations}
This means that in terms of the radial variable $\rho=\frac{1}{s}$, all deviation vector components grow linearly for $\rho\to\infty$
\begin{subequations}\label{largerinf}
\begin{equation}
\xi^{r}(r)=B_r\left(\rho-\frac{3\ln(\rho)}{E^2-1}\right)+O\left(1\right),
\end{equation}
\begin{equation}
\xi^{\theta}(\rho)=B_\theta\rho+O\left(1\right),
\end{equation}
\begin{equation}
\xi^{\phi}(\rho)=B_\phi\rho+O\left(1\right).
\end{equation}
\end{subequations}
Result (\ref{largerinf}) qualitatively coincides with (\ref{largersing}). This is to be expected because the Schwarzschild metric is asymptotically flat, and the angular momentum $l$ in flat space does not
matter because in Minkowski space geodesics are straight lines and they can deviate from each other only linearly in the radial variable $\rho=\frac{r}{M}$.

\section{\label{Conc}Conclusion}
In this article we consider tidal properties of spherical symmetric spacetimes. In Sec.~\ref{GDE} it was possible to show an alternative way to diagonalize the geodesic deviation equation in the case of geodesics with
non-zero angular momentum. We perform specific calculations in the Schwarzschild metric, which is determined only by the black hole mass. Sec.~\ref{Tforce} demonstrates that the presence of angular momentum
does not lead to the appearance of extremum points for all the tidal force components and leaves them constant in sign.

Further, in Sec.~\ref{sol} we investigate the solutions of geodesic deviation equations in Schwarzschild spacetime. First we present the previously known \cite{SSSS} solutions (\ref{elsolr})
and (\ref{elsola}) for radial geodesics, in order to see how the angular momentum affects geodesic deviation vector behavior. Eqs.~(\ref{eqr})--(\ref{eqpol}) for $l\ne0$ have no solutions in terms of
elementary functions, therefore, we found numerical solutions and presented them at different angular momentum values in Figs. \ref{rfig}--\ref{polfig} in Sec.~\ref{numsol}. These figures demonstrate
that for all Eqs.~(\ref{eqr})--(\ref{eqpol}) point $\rho=0$ is special, which is expected because it is physical singularity. Thus, in Sec.~\ref{locsol} we consider solutions near $\rho=0$, $\rho=\infty$.
In a vicinity of zero and infinity we found solutions in the form of generalized power series using the Frobenius method because Equations (\ref{eqr})--(\ref{eqpol}) satisfy Fuchs' theorem.

From a comparison of (\ref{zerorsingr}) and (\ref{lmsolr}) we found that angular momentum presence does not change the behavior of radial geodesic deviation vector component in the vicinity of zero $\rho=0$.
Therefore, regardless of fall trajectory on the Schwarzschild black hole, the test body will experience stretching, which is described by the leading order of radial geodesic deviation vector component
$\xi^r=\frac{C}{\sqrt{r}}$. Which corresponds to Fig.~\ref{rfig}. However, the behavior of the angular geodesic deviation vector components changed as expected. So the behavior of the azimuthal geodesic
deviation vector component with angular momentum appearance changed from (\ref{zerorsingaz}) $\xi^\theta=C\sqrt{r}$ to (\ref{lmsolaz}) $\xi^\theta=Cr$. Which corresponds to Fig.~\ref{azfig}. The polar
geodesic deviation vector component also changes, the appearance of angular momentum leads to the fact that square root growth (\ref{zerorsingpol}) $\xi^\phi=C\sqrt{r}$ becomes constant (\ref{lmsolpol}). This can be
seen in Fig.~\ref{polfig}.

At very large distances from the black hole, the angular momentum presence, as expected, does not change the behavior of geodesic deviation vector components, which can be seen from the comparison
(\ref{largersing}) and (\ref{largerinf}). This is due to the fact that the Schwarzschild metric is asymptotically flat and at large distances from the gravitating mass the second order differential
Eq.~(\ref{deq}) has a zero right-hand side, which means that its solution will be a linear function of $\tau$. And far from a black hole proper time $\tau$ is proportional to radial variable
$\rho=\frac{r}{M}$ according to Eq.~(\ref{req}). So the linearity of solutions of Eq.~(\ref{deq}) in vacuum and expressions (\ref{largersing}) and (\ref{largerinf}) have the same origin.

The study of tidal effects for non-radial geodesics in general relativity can evolve towards considering more general metrics, for example, Reissner--Nordstr{\"{o}}m or Kottler metric. Also quite
interesting is the question of geodesic deviation when both a test free falling body and a black hole have angular momentum. In addition, it is interesting to investigate the geodesic deviation equation
in multidimensional spacetimes and, in particular, in the metrics of black rings. As we can see, despite the fact that the classical effects in the general relativity have been studied for a very long time,
this issue still contains many unsolved problems.
\section*{Acknowledgement}
We would like to express our gratitude to Yuri Viktorovich Pavlov for meaningful discussions and useful advice.

\end{document}